\documentclass{PoS}

\newcommand{\beqn}{\begin{eqnarray}}
\newcommand{\eeqn}{\end{eqnarray}}
\newcommand{\beqs}{\begin{subequations}}
\newcommand{\eeqs}{\end{subequations}}
\newcommand{\eq}[1]{(\ref{#1})}

\newcommand{\tr}{{\mathrm{Tr}}\,}
\newcommand{\Z}{{Z \!\!\! Z}}

\newcommand{\lr}[1]{ \left( #1 \right) }

\newcommand{\vev}[1]{ \left\langle \, #1 \, \right\rangle }

\title{Vortex liquid in magnetic-field-induced superconducting vacuum of quenched lattice QCD\thanks{
The work of P. V. B. was supported by the S. Kowalewskaja award from the Alexander von Humboldt Foundation (Germany); the work of M.N.C. was partially supported by grant No. ANR-10-JCJC-0408 HYPERMAG of Agence Nationale de la Recherche (France). The work of the Moscow group was supported by Grant "Leading Scientific Schools" No. NSh-6260.2010.2, RFBR-11-02-01227-a, Federal Special-Purpose Program "Cadres" of the Russian Ministry of Science and Education and by a grant from the FAIR-Russia Research Center.  Numerical calculations were performed at the ITEP computer systems ``Graphyn'' and ``Stakan'' (authors are much obliged to A. V. Barylov, A. A. Golubev, V. A. Kolosov, I. E. Korolko and M. M. Sokolov for the valuable help).}}

\ShortTitle{Vortex liquid in superconducting vacuum of quenched lattice QCD}

\author{V. V. Braguta\\
IHEP, Protvino, Moscow region, 142284 Russia\\
ITEP, B. Cheremushkinskaya str. 25, Moscow, 117218 Russia
}

\author{P. V. Buividovich\\
Institute of Theoretical Physics, University of Regensburg,
Universit\"atsstrasse 31, D-93053 Regensburg, Germany
}

\author{\speaker{M. N. Chernodub}\thanks{On leave from ITEP, Moscow, Russia.}\\
CNRS, Laboratoire de Math\'ematiques et Physique Th\'eorique, Universit\'e Fran\c{c}ois-Rabelais Tours, Parc de Grandmont, 37200 Tours, France\\
Department of Physics and Astronomy, University of Gent, Krijgslaan 281, S9, B-9000 Gent, Belgium
}

\author{A.~Yu.~Kotov and M. I. Polikarpov\\
ITEP, B. Cheremushkinskaya str. 25, Moscow, 117218 Russia\\
MIPT, Institutskii per. 9, Dolgoprudny, Moscow Region, 141700 Russia
}

\abstract{In the background of the strong magnetic field the vacuum is suggested to possess an electromagnetically superconducting phase characterised by the emergence of inhomogeneous quark-antiquark vector condensates which carry quantum numbers of the charged $\rho$ mesons. The $\rho$-meson condensates are inhomogeneous due to the presence of the stringlike defects (the $\rho$ vortices) which are parallel to the magnetic field (the superconducting vacuum phase is similar to the mixed Abrikosov phase of a type-II superconductor). In agreement with these expectations, we have observed the presence of the $\rho$ vortices in numerical simulations of the vacuum of the quenched two-color lattice QCD in strong magnetic field background. We have found that in the quenched QCD the $\rho$ vortices form a liquid. The transition between the usual (insulator) phase at low $B$ and the superconducting vortex liquid phase at high $B$ turns out to be very smooth, at least in the quenched QCD.}

\FullConference{Xth Quark Confinement and the Hadron Spectrum\\
                 8-12 October 2012\\
                 TUM Campus Garching, Munich, Germany}
                 
\begin{document}

\section{Introduction}

Recently it was suggested that in the magnetic field background the vacuum becomes electromagnetically superconducting if the strength of the magnetic field exceeds the critical value~\cite{ref:I,ref:II}:
\beqn
B_c = m_\rho^2/e \approx 10^{16}\,\mbox{Tesla}\,,
\label{eq:eBc}
\eeqn
where $m_\rho = 775.5\,\mbox{MeV}$ is the mass of the $\rho$ meson. The magnetic fields which are two-three times stronger than the critical value~\eq{eq:eBc} are expected to be experimentally reachable at the LHC~\cite{ref:LHC:field}.

If the strength $B \equiv |\vec B|$ of the uniform static magnetic field $\vec B = (0,0,B)$ is higher then the critical value~\eq{eq:eBc}, then the usual vacuum ground state should experience a tachyonic instability towards the emergence of a new ground state characterised by the presence of the following quark-antiquark condensates:
\beqn
\langle \bar u \gamma_1 d\rangle = \rho(x_\perp)\,,
\quad \qquad
\langle \bar u \gamma_2 d\rangle =  - i \rho(x_\perp)\,.
\label{eq:ud:cond}
\eeqn
The complex scalar field $\rho$ is a function of the (transverse) spatial coordinates $x_\perp = (x_1,x_2)$.  The quark-antiquark condensates~\eq{eq:ud:cond} carry the quantum numbers of the electrically charged $\rho$ mesons so that this phenomenon may also be interpreted as the $\rho$-meson condensation. 

Since the condensed pairs~\eq{eq:ud:cond} are electrically charged states, their condensation implies, almost automatically, the emergence of the electromagnetic superconductivity in the new vacuum state at $B>B_c$. The emerging superconducting vacuum state has quite unusual properties (for a detailed review, see Ref.~\cite{ref:review:rho}):
\begin{itemize}
\item[(i)] \underline{Anisotropy}: the ground state is a perfect conductor for the electric currents directed strictly along the magnetic field axis. In the transverse directions the superconductivity is absent.

\item[(ii)] \underline{Inhomogeneity}: the transport coefficients depend on the transverse coordinates $x_\perp$.

\item[(iii)] \underline{Absence of the Meissner effect}: the induced superconductivity cannot screen the background magnetic field due to the mentioned spatial anisotropy.

\item[(iv)] \underline{In-tandem superfluidity}: one can argue that the superconducting state \eq{eq:ud:cond} should always be accompanied by a superfluidity of the neutral $\rho^{(0)}$ mesons~\cite{ref:I,ref:III}. 

\item[(iv)] Optically, the vacuum superconductor is \underline{metamaterial} with a perfect lens properties~\cite{ref:Igor}. 

\end{itemize}

In the magnetic-field-induced vacuum superconductivity, the quark-antiquark composites~\eq{eq:ud:cond} play the same role as the Cooper pairs play in a conventional superconductor. This analogy may go even deeper:  in a very strong magnetic field a conventional type-II superconductor was suggested to enter a quantum limit of ``reentrant superconductivity'' characterised by a $p$-wave spin-triplet pairing, absence of the Meissner effect, and a superconducting flow along the magnetic field axis~\cite{ref:Tesanovic}. It is encouraging that these are exactly the features which we expect to be realised in the superconducting vacuum state at $B > B_c$.

There are indications from holographic~\cite{ref:holographic,ref:holographic:hexagonal} and numerical~\cite{ref:numerical} approaches that the $\rho$-mesons should be condensed in the strong magnetic field; see also the ongoing discussion in Refs.~\cite{ref:HY,ref:reply}.

\section{Vortex lattice ground state in the mean-field approximation}

The $\rho$-meson condensation can be characterised by a single scalar function,
\beqn
\rho(x) = \frac{1}{2} {\bar u}(x) \gamma_+ d(x)\,, \quad \qquad \gamma_+ = \gamma_1 + i \gamma_2\,,
\label{eq:rho:definition}
\eeqn
where the combination $\gamma_+$  of the Dirac matrices corresponds to the $s_z = +1$ projection of the $\rho$-meson spin onto the magnetic field axis. In the small-condensate limit, $|\rho| \ll m_\rho$, the  ground state of the $\rho$-meson condensates can be described by the following general form~\cite{ref:I,ref:II,ref:III,ref:holographic:hexagonal}:
\beqn
\rho(z) = \sum_{n \in \Z} C_n \,  \exp\Bigl\{ - \frac{\pi}{2 L_B^2} \bigl(|z|^2 + {\bar z}^2\bigr) - \pi \nu^2 n^2 + 2 \pi \nu n \frac{{\bar z}}{L_B} \Bigr\}\,,
\qquad L_B = \sqrt{\frac{2 \pi}{e B}}\,,
\label{eq:rho:z} \label{eq:h:z}
\eeqn
where $L_B$ is the magnetic length and $\nu$ is a real parameter. The solution~\eq{eq:rho:z} is similar to -- and inspired by -- the Abrikosov vortex lattice configuration which appears in a mixed state of a type-II superconductor in the magnetic field background~\cite{Abrikosov:1956sx}. 

In Eq.~\eq{eq:rho:z} the complex coefficients $C_n$ are fixed by the energy minimisation condition. It is usually assumed that the coefficients $C_n$ obey the $N$--fold symmetry, where $N$ is an integer~\cite{Abrikosov:1956sx}:
\beqn
C_{n+N} = C_n\,, \qquad N = 1,2, \dots 
\label{eq:N:fold}
\eeqn
so that the condensate~\eq{eq:rho:z} has a periodic structure in the transverse $(x_1,x_2)$ plane. A condensate of the form~\eq{eq:rho:z} possesses an infinite set of zeros, which mark the centres of the topological string defects called the $\rho$ vortices~\cite{ref:I}. Similarly to the Abrikosov vortices in conventional superconductors, the condensate $\rho$ acquires the phase shift $2 \pi$ as we circumvent these zeros.

\begin{figure}[!thb]
\begin{center}
\includegraphics[scale=0.2,clip=false]{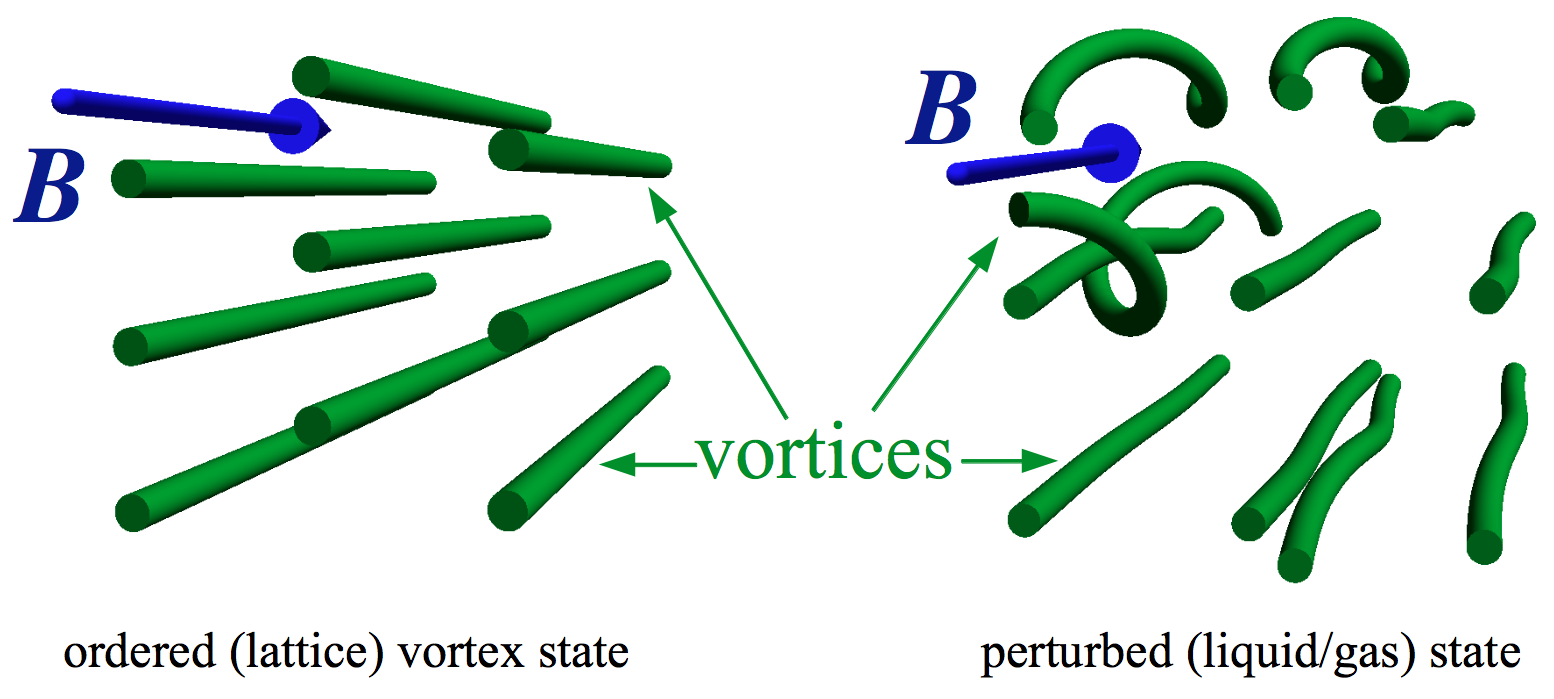}
\end{center}
\vskip -5mm
\caption{(left) The equilateral triangular vortex lattice in the mean-field approximation to the ground state of QCD and (right) its suggested melting due to the presence of quantum and/or thermal fluctuations.}
\label{fig:melting:lattice}
\end{figure}

The simplest configuration with $N = 1$ (all $C_n$'s are equal) corresponds to the square vortex lattice. However, the true mean-field ground state is given by an equilateral triangular vortex lattice (which is sometimes also called ``hexagonal lattice'') with the following set of parameters~\cite{ref:III,ref:holographic:hexagonal,Abrikosov:1956sx}:
\beqn
N=2\,,\qquad C_1 = \pm i C_0\,, \qquad \nu = \sqrt[4]{3}/\sqrt{2} \approx 0.9306\,. \qquad 
\label{eq:N2:GL}
\eeqn
Thus, in the mean field approximation, the $\rho$-meson vortex state is similar to the true ground state of the Abrikosov lattice in a type-II superconductor~\cite{Abrikosov:1956sx}; see Fig.~\ref{fig:melting:lattice}~(left) for an illustration.

The analytic results of Ref.~\cite{ref:III} suggest that the $\rho$ vortices are weakly interacting with each other in the vortex lattice state. This fact means that the presence of thermal and/or quantum fluctuations may either melt the vortex lattice state to a liquid vortex state or even evaporate it by forming a vortex gas (such phenomena are known to happen with the Abrikosov vortex lattices in conventional superconductors~\cite{ref:type-II:Review}). The vortex liquid state is a superconducting state characterised by the broken $U(1)_{\mathrm{e.m.}}$ electromagnetic gauge symmetry while the vortex gas state is, generally, a normal (non-superconducting) state with the unbroken $U(1)_{\mathrm{e.m.}}$ group. In the context of the solid state physics, the vortex lattice-liquid-gas phase diagram for magnetic-field-induced reentrant superconductivity was discussed in Ref.~\cite{ref:Tesanovic:phase}.

\section{Numerical simulations}

In our numerical setup we basically follow Ref.~\cite{ref:numerical}. We use lattice Monte-Carlo simulations of $SU(2)$ Yang-Mills lattice gauge theory. The quark fields are introduced by the overlap lattice Dirac operator $\mathcal{D}$ with exact chiral symmetry \cite{Neuberger:98:1}, and -- due to the presence of the magnetic field -- with twisted spatial boundary conditions~\cite{Wiese:08:1}. The quarks are treated in the quenched approximation so that the vacuum quark loops are absent in our approach. We have studied $18^4$ and $19^4$ lattices in a wide range of the magnetic field strengths, $eB= (0 \dots 2.14)\, \mbox{GeV}^2$ with lattice spacings $a\approx 0.11\,\mbox{fm}$. We have used 20 gauge configurations per each value of the magnetic field. 

An explicit manifestation of the superconducting phase would be the presence of the condensate~\eq{eq:ud:cond}. Unfortunately, the observable~\eq{eq:ud:cond} cannot be computed directly in our approach, while it can be accessed via the following simplest $\rho$-meson correlator:
\beqn
\phi(x) \equiv \phi(x;A,B) = \vev{\rho^\dagger(0) \rho(x)}_{A,B} \equiv  \tr\lr{\frac{1}{\mathcal{D}_u(A,B) + m} \, \gamma_{\mu} \, \frac{1}{\mathcal{D}_d(A,B) + m} \, \gamma_{\nu}}\,,
\label{eq:G:pm} \label{eq:phi}
\eeqn
where the $\rho$ meson field is defined by Eq.~\eq{eq:rho:definition}. The subscripts in Eq.~\eq{eq:G:pm} indicate that the correlation function $\phi(x)$ is computed in the fixed background of both the non-Abelian gauge field $A$ and the Abelian magnetic field $B$. Equation~\eq{eq:G:pm} represents this vector correlator in terms of the (overlap) Dirac propagators in the background of both Abelian and non-Abelian gauge fields.

We notice that under the Abelian transformation from the electromagnetic gauge group $U(1)_{\mathrm{e.m.}}$ the field~\eq{eq:phi} transforms as a charged scalar field\footnote{The gauge transformation at the origin, $\varphi(x) \to e^{i e \omega(0)} \varphi(x)$, acts as a global phase which is not essential for our interpretation of the effective field $\phi(x)$.}: $\phi(x) \to e^{i e \omega(x)} \phi(x)$. However, the effective field $\phi(x)$ is still a two-point correlation function which falls off exponentially as the distance $x$ increases. This property is not a desired behaviour for a genuine local scalar field so that the quantity~\eq{eq:phi} cannot, strictly speaking, be associated with the $\rho$-meson field itself. 

Fortunately, we may get an insight from Ref.~\cite{Bali:1994de} where a qualitatively similar issue was encountered. In that work the chromoelectric flux tube was studied using a rectangular Wilson loop ${\cal W}$ as a source and the local energy density operator ${\cal O}$ as a probe. Although the expectation value of the Wilson loop falls off exponentially as the area of the Wilson loop grows, the energy density in the presence of the Wilson loop, given by the normalised energy ratio $\vev{{\cal O}}_{\cal W} = \vev{{\cal O} \cal W}/\vev{\cal W}$, has, generally, a non-vanishing profile as the area of the Wilson loop grows.

By analogy with Ref.~\cite{Bali:1994de},  we consider the normalised scalar energy of the $\rho$-meson field $E(x)$, the normalised electric (super)current $j_\mu(x)$ generated by the $\rho$-meson field, and the local vortex density $\upsilon(x)$ in transversal $(x,y)$ plane, respectively (we use the continuum notations in order to simplify the expressions):
\beqn
E(x) & = &\frac{|D_\mu \phi(x)|^2}{|\phi(x)|^2}\,, \qquad D_\mu = \partial_\mu - i e A_\mu\,, 
\label{eq:energy}\\
j_\mu(x) & = & \frac{\phi^*(x) {\overrightarrow D}_\mu \phi(x) - \phi^*(x) {\overleftarrow D}_\mu \phi(x)}{2 i |\phi(x)|^2}\,, 
\label{eq:current}\\
\upsilon(x) & = & {\mathrm{sing}} \, {\mathrm{arg}} \, \phi(x) \equiv \frac{\epsilon^{ab}}{2\pi} \frac{\partial}{\partial x_a} \frac{\partial}{\partial x_b} \arg \phi(x)\,, \qquad a,b=1,2\,.
\label{eq:vortex}
\eeqn
 
\begin{figure}[!thb]
\begin{center}
\vskip -2mm
\begin{tabular}{cc}
\includegraphics[scale=0.18,clip=false]{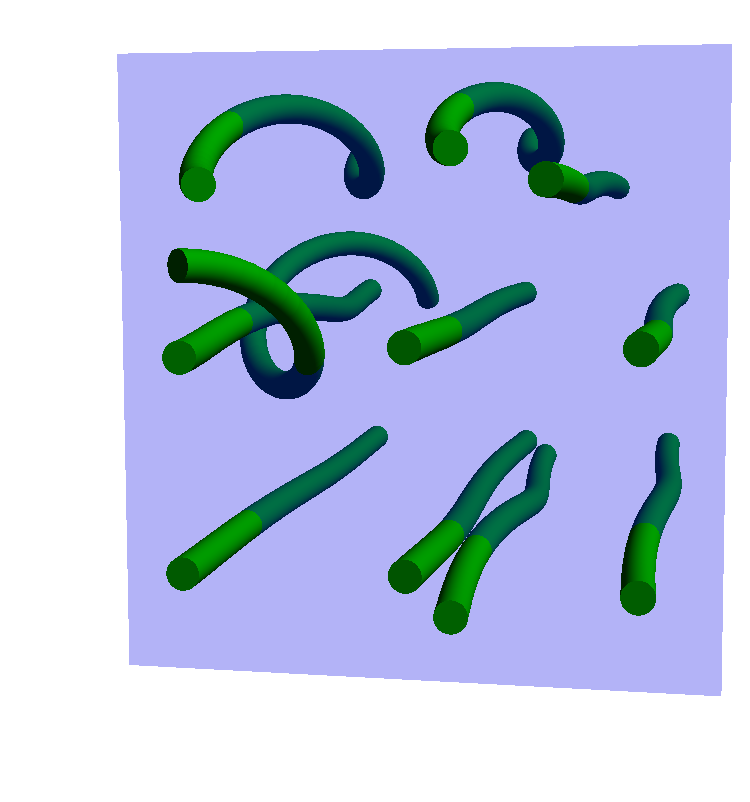} & 
\includegraphics[scale=0.18,clip=false]{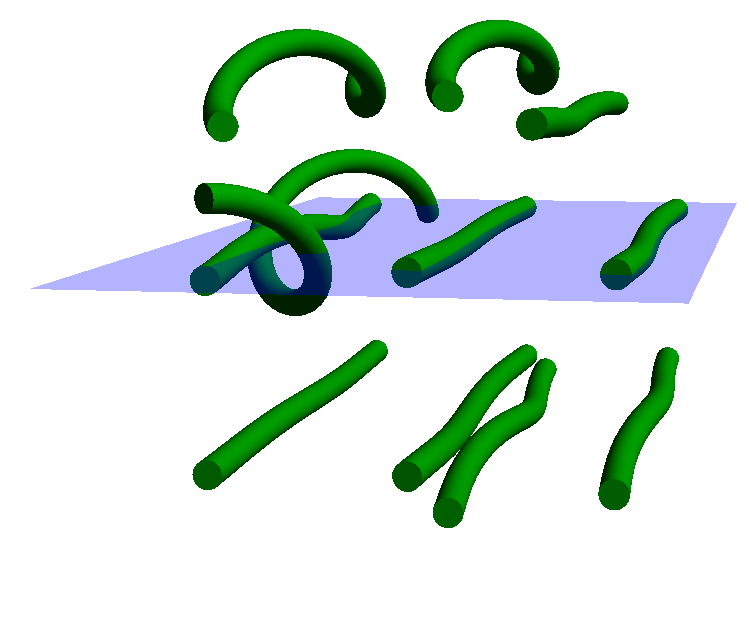} \\[-7mm]
$(x,y)$ plane & $(x,z)$ plane
\end{tabular}
\end{center}
\vskip -5mm
\caption{Two types of the ``probe'' cross-sections of the (expected) melted $\rho$-vortex lattice, Fig.~1~(right).}
\label{fig:planes}
\end{figure}
In search of signatures of the (perhaps, melted) $\rho$-vortex lattice we have studied (configuration-by-configuration) the behaviour of the normalised energy density~\eq{eq:energy} in the $(x,y)$ and $(x,z)$ planes (we remind that the magnetic field is directed along the $z$ axis);  see Fig.~\ref{fig:planes}. In the center of a physical $\rho$ vortex the energy density is higher than the energy density outside the vortex. Thus, if the physical $\rho$ vortices are formed in the (sufficiently strong) magnetic field background, than we may expect the formation of the pointlike lumps of the energy density in the $(x,y)$ plane [see Fig.~\ref{fig:planes}~(left)] and the formation of the linelike structures in the $(x,z)$ plane [see Fig.~\ref{fig:planes}~(right)].

\begin{figure}[!thb]
\begin{center}
\begin{tabular}{lr}
\includegraphics[scale=0.22,clip=false]{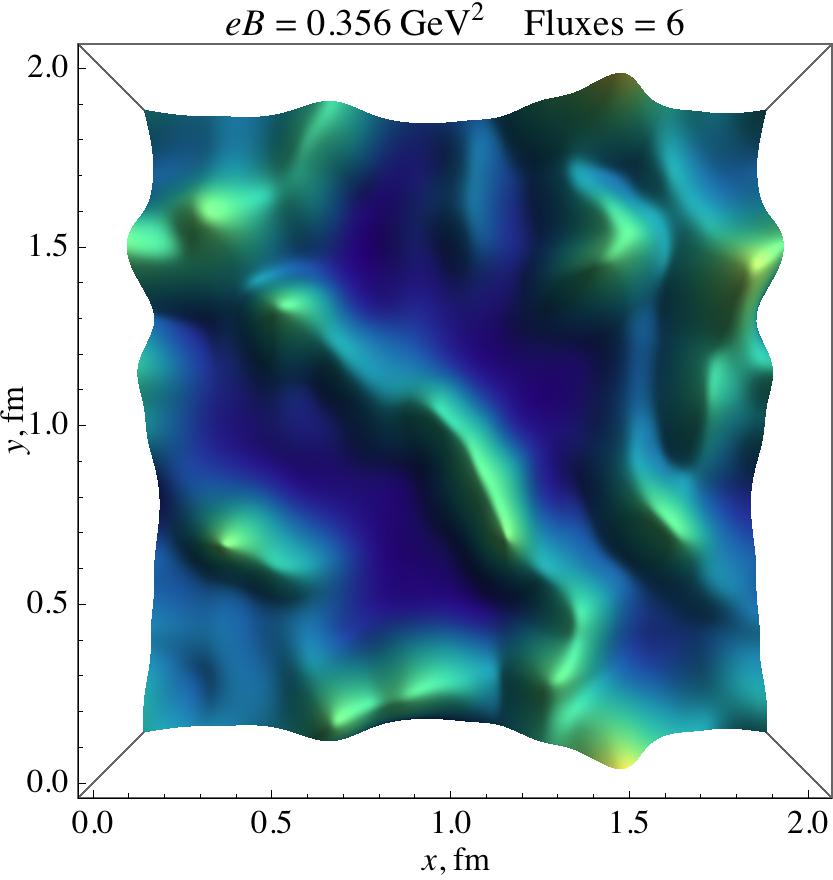} & 
\includegraphics[scale=0.22,clip=false]{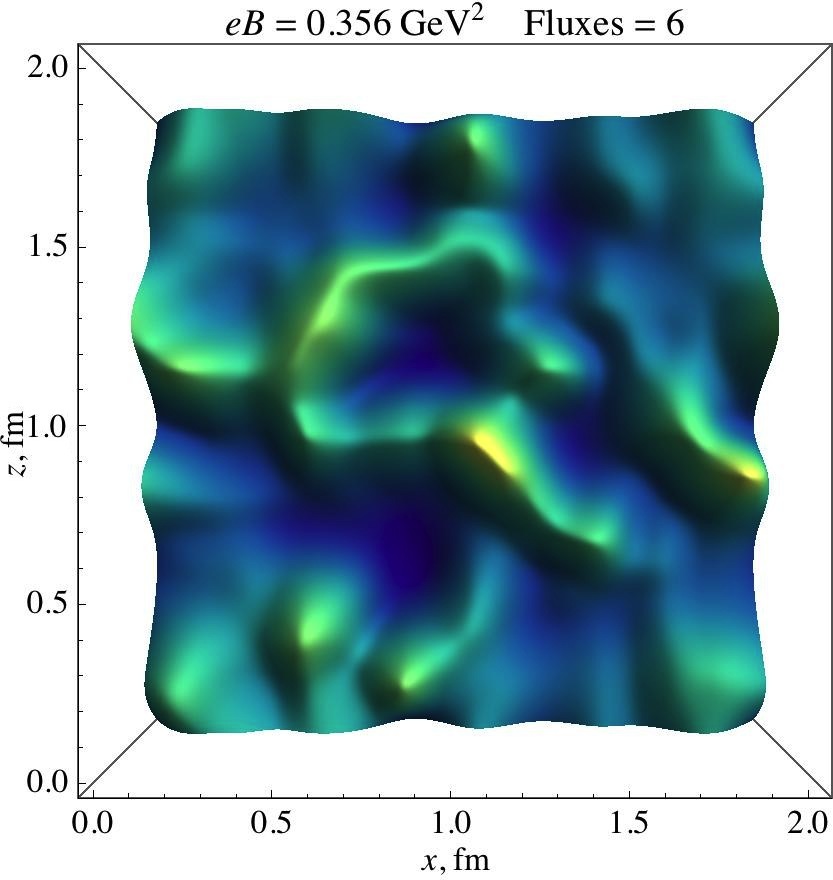} \\
\includegraphics[scale=0.22,clip=false]{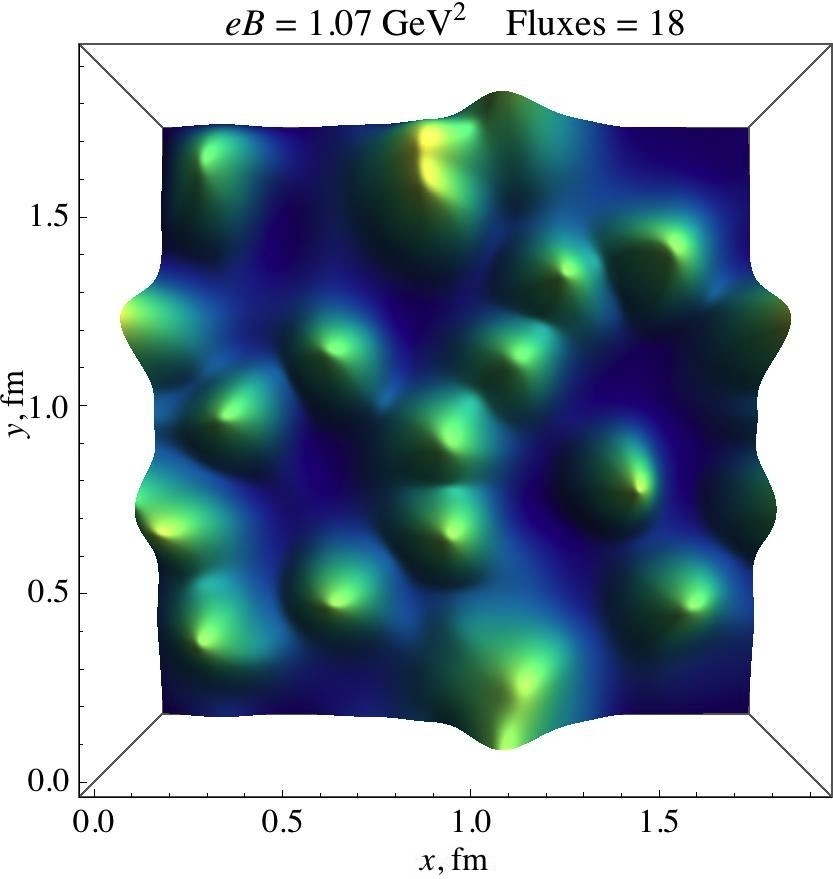} & 
\includegraphics[scale=0.22,clip=false]{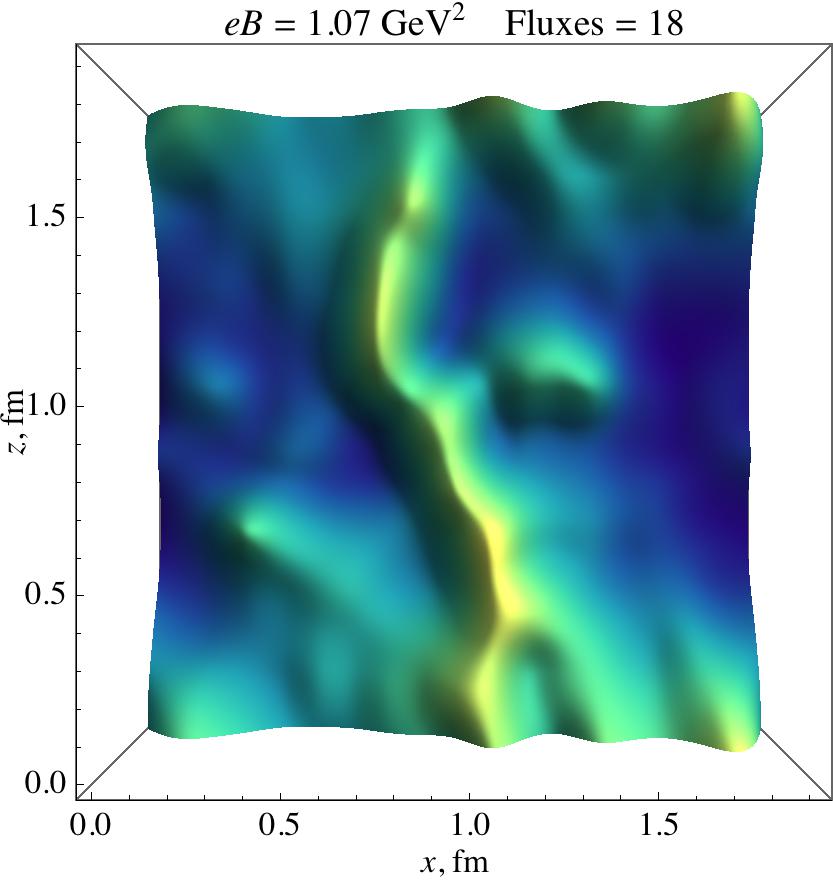} \\
\includegraphics[scale=0.22,clip=false]{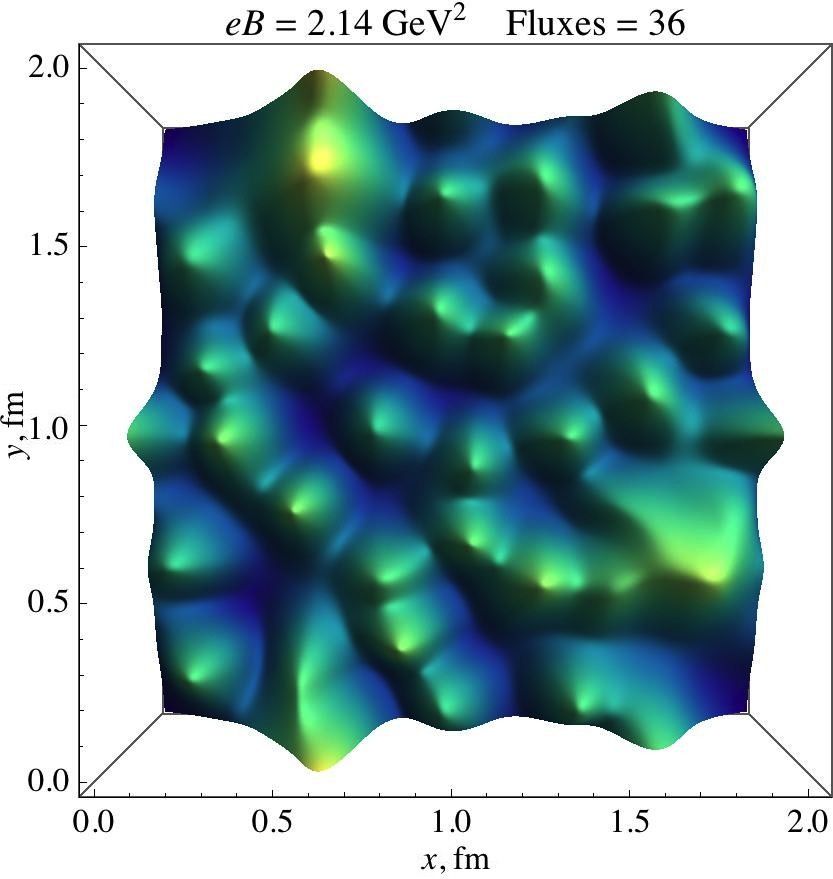} & 
\includegraphics[scale=0.22,clip=false]{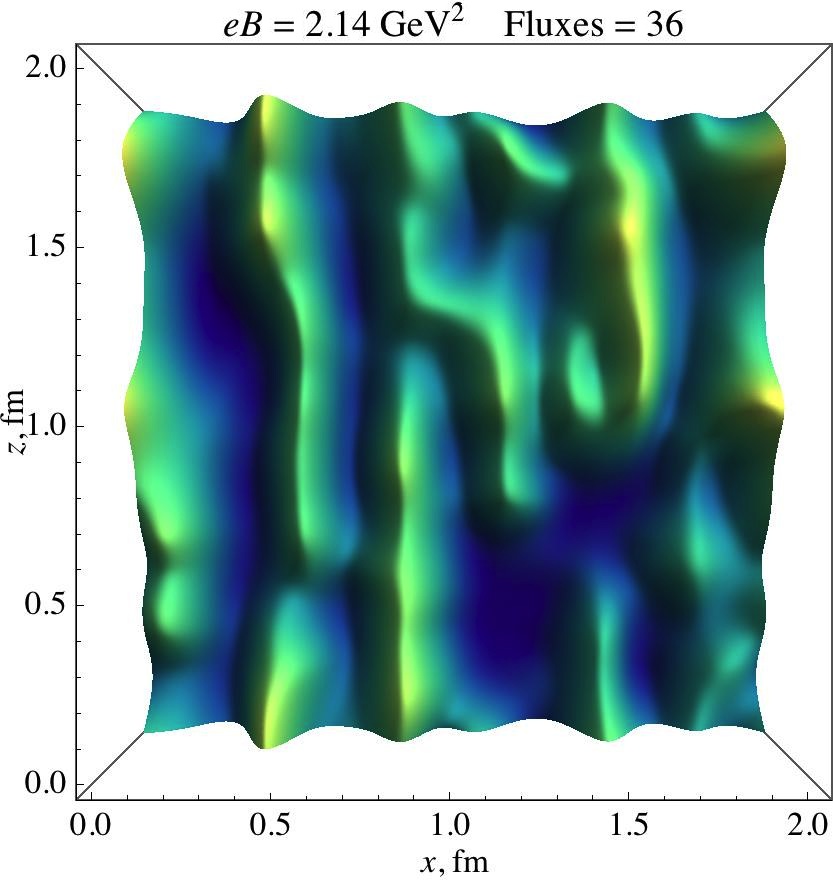}
\end{tabular}
\end{center}
\vskip -5mm
\caption{Typical behaviour of the energy density in the $(x,y)$ planes (the left column) and in the $(x,z)$ planes (the right column) for 
$eB = 0.356\,\mbox{GeV}^2$ (the upper panel),
$eB = 1.07\,\mbox{GeV}^2$ (the middle panel) and
$eB = 2.14\,\mbox{GeV}^2$ (the lower panel). The number of elementary fluxes is $n = eB L_x L_y/(2\pi)$. In the left column the magnetic field is perpendicular to the page and in the right column the magnetic field is directed vertically.}
\label{fig:energy:examples}
\end{figure}

Typical examples of the behaviour of the energy density in the $(x,y)$ and $(x,z)$ planes are shown for weak ($eB = 0.356\,\mbox{GeV}^2$), moderate ($eB = 1.07\,\mbox{GeV}^2$) and high ($eB = 2.14\,\mbox{GeV}^2$) magnetic fields in Fig.~\ref{fig:energy:examples}. In accordance with our qualitative expectations, at low magnetic field the vortex lattice is not formed. At the moderate magnetic field the formation of a coherent vortex structure is seen while the vortices are not strictly ordered in the transversal plane and they are not quite parallel to the magnetic field. At higher magnetic field the physical picture is visually consistent with the presence of a melted lattice (liquid) of the $\rho$ vortices; see Fig.~\ref{fig:melting:lattice}~(right).

\begin{figure}[!thb]
\begin{center}
\includegraphics[scale=0.21,clip=false]{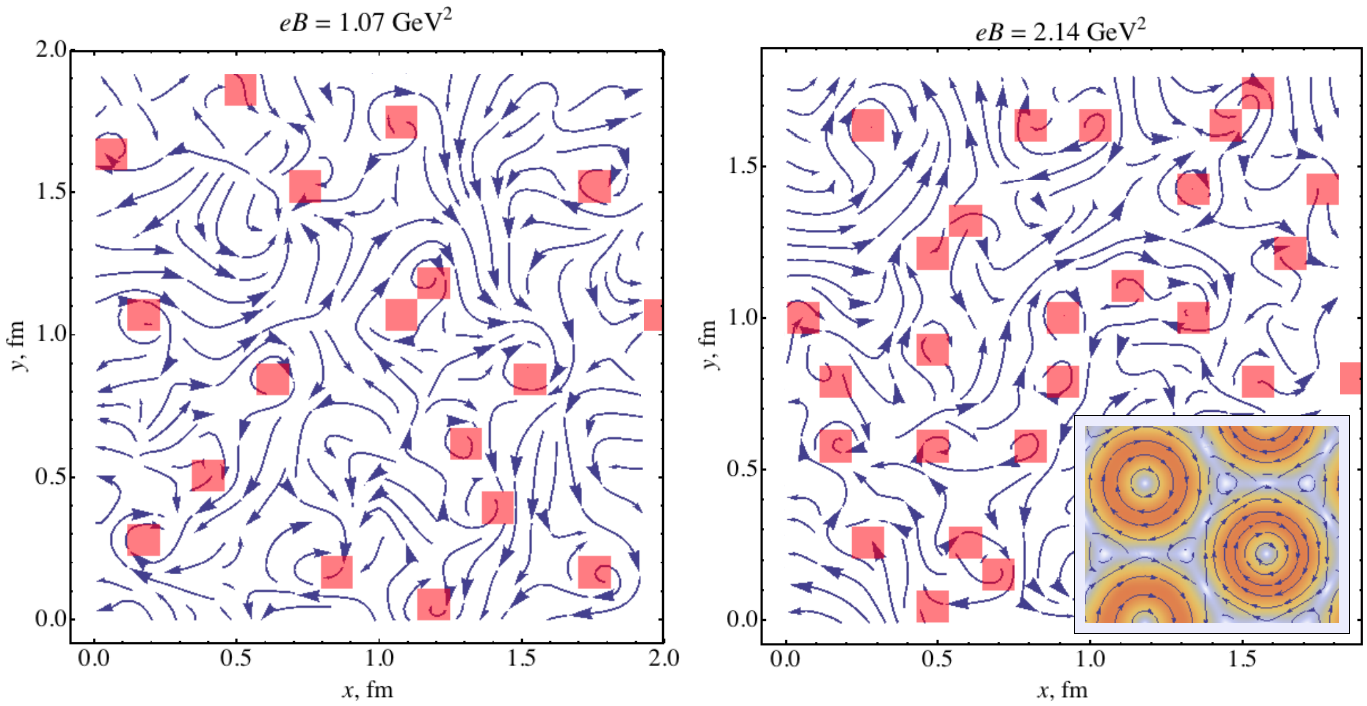}
\end{center}
\vskip -7mm
\caption{Examples of the superconducting currents (the blue lines), Eq.~(1.6), around the vortices (the red squares), Eq.~(2.7), in the $(x,y)$ planes at $eB = 1.07\, \mbox{GeV}^2$ (left) and $eB = 2.14\, \mbox{GeV}^2$ (right). According to the analytical expectations~\cite{ref:I,ref:III} the currents should encircle the $\rho$ vortices in the clockwise direction (an example of the mean-field solution, Eqs.~(1.4), (1.6), is shown in the inset of the right figure, from Ref.~\cite{ref:III}).} 
\label{fig:curl}
\end{figure}

The peaks in the energy density~\eq{eq:energy} are correlated with the $\rho$ vortex positions~\eq{eq:vortex} and that the $\rho$ vortices are encircled by the supercurrents~\eq{eq:current}. The latter feature is shown in Fig.~\ref{fig:curl}. Thus, the numerically observed vortices do indeed bear the essential features of the physical vortices.

\begin{figure}[!thb]
\begin{center}
\begin{tabular}{lr}
\includegraphics[scale=0.24,clip=false]{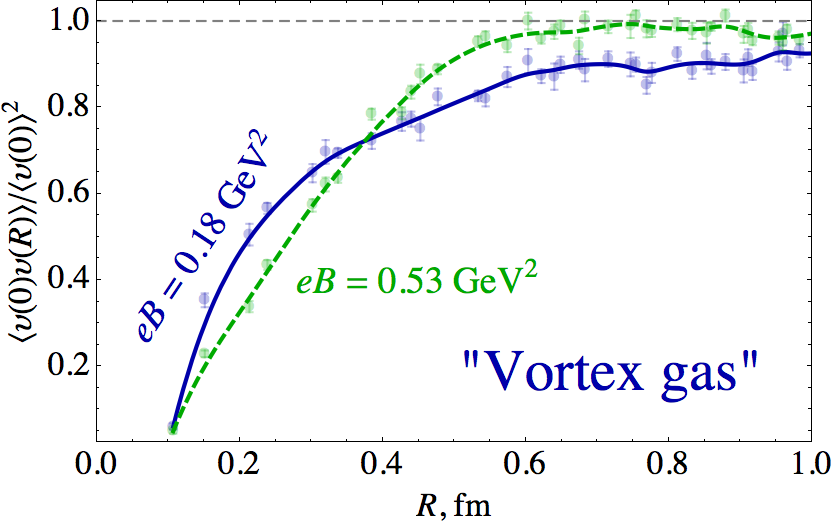} &
\includegraphics[scale=0.24,clip=false]{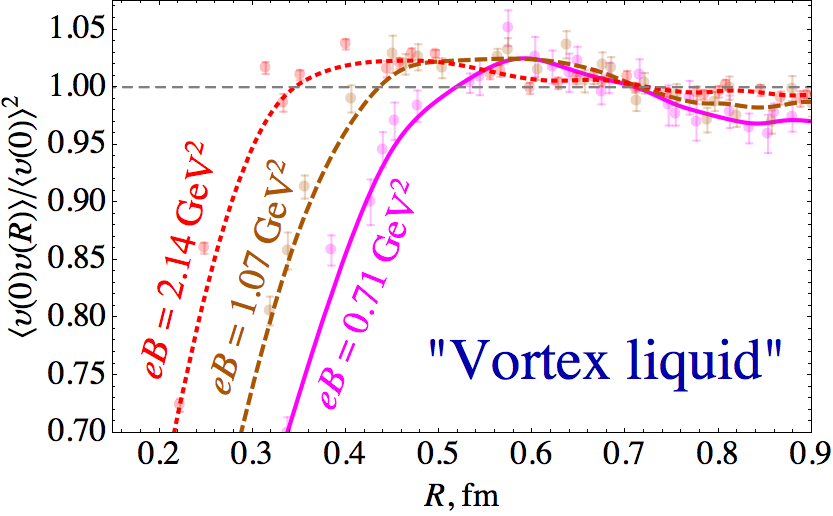}
\end{tabular}
\end{center}
\vskip -7mm
\caption{The normalised vortex-vortex correlation function $\vev{\upsilon(0) \upsilon(R)}/\vev{\upsilon(0)}^2$ in the $(x,y)$ plane. The monotonic (non monotonic) behaviour of the correlator signals the presence of the gas (liquid) vortex state at low (high) values of the magnetic field $B$ as shown in the left (right) plot.}
\label{fig:correlations}
\end{figure}

The nature of the $\rho$-vortex state may be characterised by the normalised vortex-vortex correlation function $\vev{\upsilon(0) \upsilon(R)}/\vev{\upsilon(0)}^2$, where the $\rho$ vortex density is given in Eq.~\eq{eq:vortex}. At low magnetic fields this function is a monotonically rising function of the inter-vortex distance~$R$, Fig.~\ref{fig:correlations}~(left), implying that the $\rho$ vortices constitute a (nonsuperconducting) gas.

\clearpage
\clearpage

At higher magnetic fields a non-monotonic behaviour manifests itself via the formation of a wide maximum at intermediate distances; see Fig.~\ref{fig:correlations}~(right). The appearance of the peak indicates the presence of the anticipated superconducting $\rho$-vortex liquid. 

The melting of the vortex lattice in quenched QCD may make it difficult to observe the suggested $\rho$ vortex condensation using the standard numerical tools. Indeed, in the vortex lattice state [described by Eqs.~\eq{eq:rho:z} and Eq.~\eq{eq:N2:GL}] the phase of the $\rho$-meson field changes by $2\pi$ around each vortex so that the $\rho$-field is an oscillatory function of the transverse coordinates $x$ and $y$. Thus, in the superconducting state at $B > B_c$, the space-averaged (bulk) condensate is always zero, $\left\langle \rho(x) \right\rangle_{\mathrm{bulk}} \equiv 0$, despite the fact that the local $\rho$-meson condensate is large and the ground state is a superconductor (the same is true for the Abrikosov mixed state in a type-II superconductor~\cite{Abrikosov:1956sx}). 

However, if the vortices were strictly straight, then the $\rho$-meson condensation could in principle still be determined by studying a long-distance limit (taken along the straight vortex worldsheets) of the correlation function~\eq{eq:G:pm} averaged over gluon fields. In this straight-vortex case the oscillating phase would not contribute to the correlation function so that long-distance correlator should generally be nonzero, $\lim\limits_{x^\| \to\infty} \left\langle \rho(0) \rho(x^\|) \right\rangle \sim |\vev{\rho(0)}|^2$ with $x^\| = (0,0,x,t)$. However, in the liquid vortex phase the vortex worldsheets are not flat surfaces so that the vortex wobbling may, in general, add large phase fluctuations to the correlator of the $\rho$-meson field, hence $\lim\limits_{x^\|\to\infty} \left\langle \rho(0) \rho(x^\|) \right\rangle \equiv 0$ in this superconducting state. Thus, 
one may encounter a technical difficulty in determination of the exact $\rho$-meson condensate in the liquid state by using the $\rho$-field correlators.

\section{Conclusions}

We have numerically observed the formation of the $\rho$-vortex liquid in the vacuum of quenched two-color QCD in strong magnetic field background. The vortex liquid phase is an electromagnetically superconducting phase characterised by the inhomogeneous  order parameter (the $\rho$ meson condensate), similarly to the mixed (Abrikosov) phase of an ordinary type-II superconductor. We argue that in this phase the calculation of the (highly inhomogeneous) $\rho$-meson condensate by using the standard methods should be taken with care. The transition between the usual (insulator) phase at low $B$ and the superconducting vortex liquid phase at high $B$ turns out to be very smooth.

\end{document}